\documentclass[journal]{IEEEtran}

\usepackage[dvipsnames]{xcolor}
\usepackage[colorlinks=true,bookmarks=false,linkcolor=NavyBlue,urlcolor=NavyBlue,citecolor=NavyBlue,breaklinks]{hyperref}
\usepackage{amsmath}
\usepackage{amsfonts}
\usepackage{graphicx}
\usepackage[italicdiff]{physics}
\usepackage{mathtools}
\usepackage{setspace}

\newcommand{\intall}{\int_{-\infty}^{\infty}}
\newcommand{\avg}[1]{\langle#1\rangle}
\newcommand{\Avg}[1]{\left\langle#1\right\rangle}

\newcommand{\floor}[1]{\left\lfloor#1\right\rfloor}
\newcommand{\ceil}[1]{\left\lceil#1\right\rceil}

\newcommand{\iverson}[1]{1_{#1}}
\newcommand{\new}[1]{#1}

\DeclareMathOperator{\sinc}{sinc}

\DeclareMathOperator{\He}{He}
\DeclareMathOperator{\expect}{\mathbb E}
\DeclareMathOperator{\vari}{\mathbb V}

\begin{document}

\title{Efficient superoscillation measurement for incoherent
  optical imaging}

\author{Mankei Tsang \thanks{This research is supported by the National
    Research Foundation (NRF) Singapore, under its Quantum Engineering
    Programme (Award~QEP-P7).  Mankei Tsang is with the Department
    of Electrical and Computer Engineering, National University of
    Singapore, 4 Engineering Drive 3, Singapore 117583 and the
    Department of Physics, National University of Singapore, 2 Science
    Drive 3, Singapore 117551 (email:
    \href{mailto:mankei@nus.edu.sg}{mankei@nus.edu.sg}, website:
    \url{https://blog.nus.edu.sg/mankei/}).}  }

\maketitle

\begin{abstract}
  I propose a superoscillation measurement method for subdiffraction
  incoherent optical sources, with potential applications in
  astronomy, remote sensing, fluorescence microscopy, and
  spectroscopy.  The proposal, based on coherent optical processing,
  can capture all the light on the aperture in principle, perform
  better than direct imaging on statistical terms, and approach the
  fundamental quantum limit.
\end{abstract}

\section{\label{sec_intro}Introduction}
Improving the resolution of an optical imaging system beyond the
diffraction limit has been a dream of opticians and a subject of
intense research for centuries \cite{villiers}. Superresolution
research has two elephants in the room, however: signal-to-noise ratio
(SNR) and competition with computational techniques, such as
deconvolution \cite{villiers}. Research on superoscillation, which has
received renewed interest in recent years \cite{berry19}, is no
exception. Existing techniques rely on masking parts of the aperture,
meaning that a significant portion of the light is lost. It is also
unclear whether the enhancement is superior to that obtainable by
digital image processing together with full-aperture direct imaging,
which can capture as much light as the aperture allows.

Building on the recent work on quantum-inspired superresolution
\cite{tnl,tsang19a}, here I show how a superoscillation measurement
can be implemented for incoherent sources without losing any of the
light in principle. I also show that the resultant error is much lower
than that obtainable by direct imaging with image processing for
subdiffraction objects. In fact, the error is close to the quantum
limit according to the quantum Cram\'er-Rao bound (QCRB)
\cite{helstrom}, so there is little room for further improvement, as
far as passive imaging is concerned. The proposed method works in the
far field, requires only known optical technologies and materials, and
has diverse potential applications in astronomy, remote sensing,
fluorescence microscopy, and spectroscopy.

\new{The key insight of this work is that the Fourier coefficients of
  an object intensity function, in terms of a basis of
  superoscillatory functions, can be constructed from the object
  moments. As the moments can be measured efficiently via spatial-mode
  demultiplexing (SPADE)
  \cite{tsang17,tsang18a,tsang19,tsang19b,tsang21a,yang16,zhou19,bonsma19,liang21a},
  the Fourier coefficients can also be measured efficiently. This work
  hence serves as a bridge between the traditional treatment of
  superresolution based on Fourier analysis and the nascent field of
  quantum-inspired incoherent imaging, which has so far focused on
  special parametric models or moment estimation only.}

The theory here turns out to share some similarities with the
singular-system approach pioneered by Slepian and Bertero
\cite{villiers,slepian83,bertero} as well. This connection is not
surprising, considering that the singular-system approach is
intimately related to superoscillation \cite{villiers}, but it is
still satisfying to have a unified picture here. The important
difference of the proposal here from the method proposed by Bertero
and coworkers for incoherent imaging \cite{bertero,walker93} is that
the former involves coherent optical processing, whereas the latter
processes the image-plane intensity only and is subject to the same
limits as those for direct imaging.

The statistical and quantum analysis here is novel in the context of
superoscillation research \cite{berry19}.  It is noteworthy that
Kolobov and coworkers performed a similar kind of analysis for
coherent imaging in the context of the singular-system approach
\cite{kolobov_fabre,beskrovnyy05,beskrovny08,piche}, but they did not
study the incoherent case, which is arguably more important in
optics. Besides the obvious necessity of considering incoherent
sources for astronomy and remote sensing, it is also necessary to use
incoherent fluorophores in biological microscopy to provide
protein-specific tagging and contrast
\cite{pawley,moerner15,betzig15,hell15}; label-free optical methods
are unable to provide such contrast and also cannot compete with
electron microscopy in many applications. Compared with existing
superresolution techniques in fluorescence microscopy that manipulates
the fluorophore emission \cite{moerner15,betzig15,hell15}, far-field
methods that extract more information from the light may complement or
supersede them by covering for their shortcomings, such as slow speed
and phototoxicity.

To be sure, the achievable resolution enhancement is still severely
limited by the photon budget and the object size. It is in the sense
of making almost the best use of the incoming photons that I claim the
measurement to be efficient.

\new{
\section{\label{sec_review}Review of superresolution and superoscillation}
To set the stage, I first review the concepts of superresolution and
superoscillation that are relevant to this work.  This section is
mostly based on Ref.~\cite{villiers}.

Let $\mathcal F$ be the input function space, where each element is a
function $F:D \to E$ that determines the optical fields emitted by an
object on the object plane. Define similarly an output function space
$\mathcal F'$, which consists of output signals $f:D'\to E'$ that can
be measured on the image plane. Let $\Pi:\mathcal F \to \mathcal F'$
be a linear operator that models the imaging system. Assume that $\Pi$
is a low-pass filter with a bandwidth that can be normalized to a
dimensionless number.  In optical imaging, a fundamental mechanism of
the low-pass filtering is the diffraction limit due to a finite
numerical aperture \cite{goodman}.  For example, $\Pi$ may be
represented by the convolution
\begin{align}
f(x) &= \int_D h(x-X) F(X) d^m X,
&
F &\in \mathcal F,
&
f &\in \mathcal F'.
\end{align}
Common examples of the kernel $h$ include the Gaussian
$h(x) \propto \exp(-x^2/2)$ and the sinc function
$h(x) \propto \sinc x$, defined as $\sinc x \equiv (\sin x)/x$ if
$x \neq 0$ and $\sinc 0 \equiv 1$.  If the domain $D$ of the input
function is assumed to be the whole Euclidean space $\mathbb R^m$,
then elementary Fourier analysis shows that the frequency components
of the input function outside the filter bandwidth are blocked or
severely attenuated and cannot be observed from the output, especially
if there is noise.

The overarching principle of superresolution imaging is to restrict
the input function space $\mathcal F$, either by assumption or by
experimental control, such that the low-pass filter is less harsh on
the smaller function space and it becomes possible to measure certain
object features with feature sizes somewhat smaller than the inverse
of the filter bandwidth.

For convenience, I normalize the object-plane coordinate with respect
to the filter bandwidth. Then, roughly speaking, a feature size is
considered superresolution if it is somewhat smaller than $1$. For
example, if the object can be assumed to consist of a finite number of
point sources, then there are only a finite number of unknown scalar
parameters, and it becomes possible to estimate the source positions
with superresolution precision by image processing, given a high
enough SNR \cite{schiebinger}. Studies in quantum metrology have also
shown that judicious measurements can offer much better SNRs when
observing certain superresolution features of a parametric model, such
as the positions of a few point sources
\cite{tnl,tsang19a,helstrom,bisketzi19} or the size of an object with
a given shape \cite{tsang17,dutton19,krovi22}.

To deal with more general objects, there is a need for a parameter
space that is high-dimensional or even infinite-dimensional.  When the
parameter space is infinite dimensional, the estimation problem is
called a semiparametric problem \cite{bickel93}.  One of the first
superresolution ideas that works under the semiparametric setting, as
pioneered by Slepian, Bertero, and coworkers \cite{slepian83,bertero},
is to assume that the domain $D$ of the input functions is bounded,
such as $[-\Delta,\Delta]$ with $0 < \Delta < \infty$, so that the
low-pass filter is softer on those functions. To be precise, assume
that $\mathcal F$ and $\mathcal F'$ are Hilbert spaces with the inner
product denoted by $\avg{\cdot,\cdot}$.  If $\Pi$ is compact, it can
be expressed as the singular-value decomposition (SVD)
\begin{align}
\Pi F &= \sum_\mu s_\mu b_\mu' \Avg{b_\mu,F},
&
F &\in \mathcal F,
\label{SVD}
\end{align}
where $\{s_\mu\}$ are a set of positive scalars called singular values
and $\{b_\mu\}$ and $\{b_\mu'\}$ are called input and output singular
functions, which are orthonormal subsets of $\mathcal F$ and
$\mathcal F'$, respectively.  If $D$ is bounded, $\{b_\mu\}$ may be
able to span $L^2(D)$, the Hilbert space of all square-integrable
functions on $D$, meaning that all the functions in
$\mathcal F = L^2(D)$ can survive the filter. The input signal can be
decomposed in a (generalized) Fourier series as
\begin{align}
F(X) &= \sum_\mu \beta_\mu b_\mu(X),
&
\beta_\mu &= \Avg{b_\mu, F}.
\label{fourier}
\end{align}
Then, in the noiseless case, each Fourier coefficient $\beta_\mu$ can
be retrieved from the output function $f = \Pi F$
by the linear filter
\begin{align}
\beta_\mu &= \frac{1}{s_\mu} \Avg{b_\mu',f}.
\label{inverse}
\end{align} 
If noise is present in the output, the error of estimating each
$\beta_\mu$ depends on the magnitude of $s_\mu$ relative to the noise
level.

For many imaging problems in one dimension ($D \subseteq \mathbb R$),
each singular function with index $\mu \in \mathbb N_0$ is an
oscillatory function with $\mu$ zeros within the domain, so the
feature size of each singular function is roughly $\Delta/(\mu+1)$.  A
well known example is the prolate spheroidal functions for both
$\{b_\mu\}$ and $\{b_\mu'\}$ when $D = [-\Delta,\Delta]$ and
$h(x) \propto \sinc x$ \cite{slepian83}. While the singular values
stay nonzero for all $\mu \in \mathbb N_0$, they decrease rapidly with
increasing $\mu$ when $\Delta/(\mu+1)$ becomes significantly smaller
than $1$, so superresolution is severely limited by the SNR in
practice.

Kolobov and coworkers generalized this singular-system approach for
coherent imaging in quantum optics by assuming that the input and
output functions are the mean fields of certain quantum states, to be
measured by homodyne detection in terms of the singular modes
\cite{kolobov_fabre,beskrovnyy05,beskrovny08,piche}. They also
proposed the use of a multimode squeezed state to reduce the noise,
but it remains an open question how one can make an object emit the
desired squeezed state in a real application.

More recently, Refs.~\cite{bearne21,matlin22} have studied numerically
the application of SPADE to incoherent imaging for general objects.
Much work remains to be done, however, to understand and prove the
advantage on a theoretical level.

The related idea of superdirectivity aims to produce an output
electromagnetic field with superresolution features by controlling the
input \cite{francia52,huang_zheludev}. Provided that the output
singular functions $\{b_\mu'\}$ of an electromagnetic system are also
oscillatory within a bounded domain $D'$, superresolution in the
output may be achieved by an input function $F$ with Fourier
coefficients $\{\beta_\mu\}$ that are designed to compensate for the
decay of the singular values $\{s_\mu\}$. Given the close relation
between focusing and imaging \cite{goodman}, the superdirectivity
concept can be exploited to give the point-spread function of an
optical imaging system certain superresolution features by masking
parts of the aperture \cite{francia52,huang_zheludev}. It is not at
all clear, however, whether such a lossy imaging system can compete
with other superresolution methods on statistical terms when noise is
present. 

The singular-system theory can be related to the phenomenon of
superoscillation by considering the singular functions $\{b_\mu\}$ and
$\{b_\mu'\}$ outside their bounded domains \cite{villiers}. As $\Pi$
is a low-pass filter when the function domains are taken to be the
whole Euclidean space $\mathbb R^m$, $\{b_\mu\}$ and $\{b_\mu'\}$
should also be bandlimited functions over $\mathbb R^m$, meaning that
their fast oscillations within the bounded domains are necessarily
accompanied by large sidelobes outside, as required for
superoscillatory functions \cite{villiers,berry19}.

For the arguably more important case of incoherent imaging, Bertero
and coworkers \cite{bertero} considered the classical direct-imaging model
\begin{align}
f(x) &= \int_D |\psi(x-X)|^2 F(X)  d^m X,
\label{direct}
\end{align}
where $F$ is the intensity of an object emitting spatially incoherent
light, $\psi$ is the point-spread function of a diffraction-limited
imaging system with respect to the optical fields, and $f$ is the
intensity on the image plane \cite{goodman}. Assuming that $D$ is
bounded or, more generally, that the object is illuminated by a
focused beam with intensity $I(X)$ such that
$f(x) = \int |\psi(x-X)|^2 F(X) I(X) d^m X$ in a confocal microscope,
Bertero and coworkers proposed the processing of $f$ based on the
singular-system theory \cite{bertero}, via the use of intensity masks
for example \cite{walker93}. The proposal looks similar to the
image-scanning microscopy proposed by Sheppard \cite{sheppard88} and
M\"uller and Enderlein \cite{mueller10}, which also involves the
processing of the image $f$ to achieve modest superresolution,
although the latter does not explicitly refer to the singular-system
theory.

It is apparent from the preceding discussion that the SNR is a major
concern for superresolution imaging, so it is desirable to minimize
any unnecessary loss in the imaging system.  It is also unclear which
superresolution method performs the best statistically. This work
proposes a measurement scheme that addresses these two issues: the
scheme does not introduce any loss intentionally and its statistical
performance in the presence of photon shot noise is close to the
fundamental quantum limit imposed by quantum estimation theory
\cite{tsang19a,helstrom}. 
}

\section{Key ideas}

\subsection{\label{sec_summary}Proposal summary}
I focus on one-dimensional paraxial imaging for simplicity
($D \subseteq \mathbb R$) \cite{goodman}. Let $F(X)$ be the
nonnegative object intensity function that is normalized as
\begin{align}
\intall F(X) dX = 1.
\end{align}
Assume that the object size is subdiffraction---to be specific, assume
\begin{align}
F(X) = 0 \textrm{ for }|X| > \Delta,
\label{subdiff}
\end{align}
where $0 < \Delta \ll 1$. Such objects are, of course, abundant in
astronomy, while the condition may also be enforced in microscopy by a
confocal illumination \new{with stimulated-emission depletion around a
  spot with size $\Delta$ \cite{pawley}. Apart from the assumptions
  above, $F$ is assumed to be arbitrary, meaning that the problem is
  semiparametric. The goal of the proposal is to measure the Fourier
  coefficients
\begin{align}
\beta_\mu &= \intall b_\mu(X) F(X) dX,
\label{beta}
\end{align}
where $\{b_\mu:\mu \in \mathbb N_0\}$ are a set of orthonormal
functions with respect to a certain inner product and each $b_\mu(X)$
is oscillatory with $\mu$ zeros in $[-\Delta,\Delta]$. The feature
size of each $b_\mu(X)$ function is then roughly $\Delta/(\mu+1)$,
which is much smaller than the diffraction-limited feature size on the
order of $1$. 

The goal set forth is in the same spirit as the singular-system
approach discussed in Sec.~\ref{sec_review}, except that the basis
$\{b_\mu\}$ here does not come from a SVD. Rather, each $b_\mu(X)$ is
a polynomial
\begin{align}
b_\mu(X) &= \sum_{\nu=0}^\mu B_{\mu\nu} X^\nu,
\label{polynomial}
\end{align}
such that each Fourier coefficient can be expressed as
\begin{align}
\beta_\mu &= \sum_{\nu=0}^\mu B_{\mu\nu} \theta_\nu,
\label{fourier_moment}
\end{align}
where 
\begin{align}
\theta_\nu &\equiv \intall X^\nu F(X) dX
\label{moment}
\end{align}
is an object moment. As the object moments can be estimated
efficiently by SPADE
\cite{tsang17,tsang18a,tsang19,tsang19b,tsang21a}, each Fourier
coefficient $\beta_\mu$ of order $\mu$ can be reconstructed from the
object moments $\{\theta_0,\theta_1,\dots,\theta_\mu\}$.

The connection to superoscillation comes from the fact that, if one
takes the whole $\mathbb R$ to be the domain of $F$, the
diffraction limit of the imaging system still imposes a bandwidth
limit to each $b_\mu(X)$, so the oscillations of $b_\mu(X)$ within
$[-\Delta,\Delta]$ should be accompanied by large sidelobes outside
the interval.

As explained later, the actual $b_\mu(X)$ functions implemented by the
proposed method are only approximations of the desired polynomials,
although the approximations can be quite accurate and still exhibit
the desired oscillatory behaviors when $\Delta \ll 1$.

}

\subsection{Orthogonal polynomials}
\begin{figure}[htbp!]
\centerline{\includegraphics[width=0.48\textwidth]{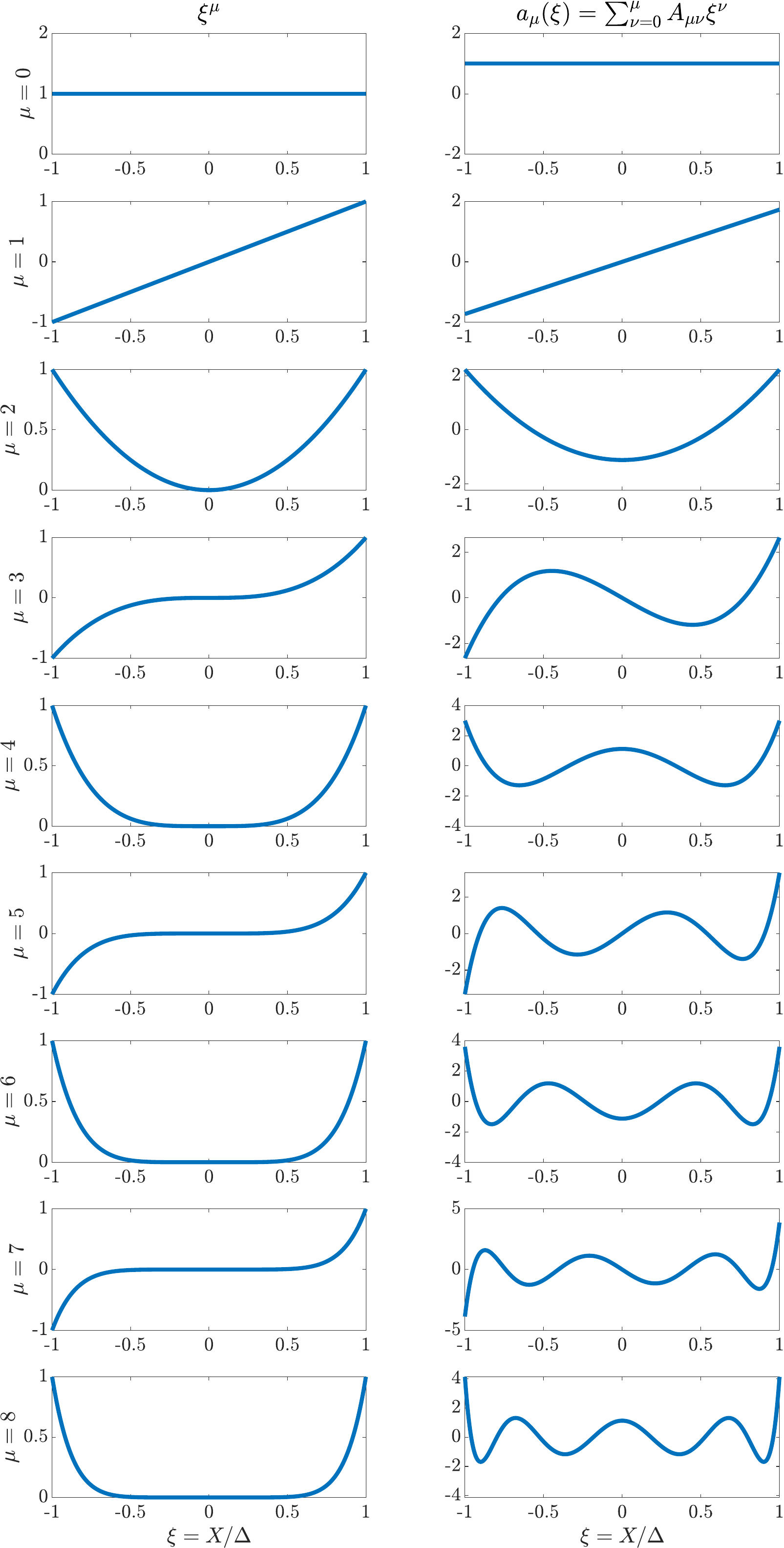}}
\caption{\label{legendre_plot}Left column: plots of monomials
  $\xi^\mu$ from $\mu = 0$ to $\mu = 8$. Right column: plots of
  the orthogonal polynomials $a_\mu(\xi)$ given by
  Eq.~(\ref{legendre}). As each orthogonal polynomial $a_\mu(\xi)$ of
  order $\mu$ can be constructed from the monomials
  $\{\xi^\nu: \nu = 0,1,\dots,\mu\}$, it follows that each Fourier
  coefficient $\beta_\mu = \intall a_\mu(X/\Delta) F(X) dX$ can be
  constructed from the moments
  $\{\theta_\nu = \intall X^\nu F(X) dX:\nu = 0,1,\dots,\mu\}$.}
\end{figure}

Before explaining the proposal in further detail, it is necessary to
introduce the concept of orthogonal polynomials first. It is helpful
to perform a further normalization by writing
\begin{align}
F(X) = \frac{1}{\Delta} W\qty(\frac{X}{\Delta}),
\label{W}
\end{align}
where $W$ is the object intensity function with a normalized width.
Suppose that $W$ can be expanded in a generalized Fourier series as
\begin{align}
W(\xi) &= \sum_{\mu=0}^\infty \beta_\mu a_\mu(\xi) R(\xi),
\\
a_\mu(\xi) &= \sum_{\nu=0}^\mu A_{\mu\nu}\xi^\nu,
\label{a}
\end{align}
where $R(\xi)$ is a reference density, such as the rectangle function
\begin{align}
R(\xi) &= \frac{\iverson{|\xi| \le 1}}{2},
\label{rect}
\\
\iverson{\textrm{proposition}} &\equiv \begin{cases}
1 & \textrm{if proposition is true},\\
0 & \textrm{otherwise},
\end{cases}
\end{align}
$\{a_\mu(\xi): \mu \in \mathbb N_0\}$ are the orthonormal polynomials
with respect to the real $R$-weighted inner product \cite{olver}
\begin{align}
\Avg{a_\mu(\xi),a_\nu(\xi)}_{R(\xi)} 
\equiv \intall a_\mu(\xi)a_\nu(\xi) R(\xi) d\xi
= \delta_{\mu\nu},
\end{align}
and each $\beta_\mu \in \mathbb R$ is a Fourier coefficient that can
also be expressed as
\begin{align}
\beta_\mu &= \intall a_\mu(\xi) W(\xi) d\xi = 
\intall a_\mu\qty(\frac{X}{\Delta}) F(X) dX.
\end{align}
The coefficient matrix $A$ in Eq.~(\ref{a}) can be derived by applying
the Gram-Schmidt procedure to the monomials
$\{1,\xi,\xi^2,\dots\}$. The procedure implies the property
\begin{align}
\Avg{a_\mu(\xi),\xi^\nu}_{R(\xi)} = 0 \textrm{ for }\mu > \nu \in \mathbb N_0,
\end{align}
which will be useful throughout this paper. The desired $B$
coefficients in Eqs.~(\ref{polynomial}) and (\ref{fourier_moment})
become
\begin{align}
B_{\mu\nu} &= \frac{A_{\mu\nu}}{\Delta^\nu}.
\end{align}
For example, for the rectangle $R$ given by Eq.~(\ref{rect}), the
orthonormal polynomials are given by
\begin{align}
a_\mu(\xi) = \sqrt{2\mu+1} P_\mu(\xi),
\label{legendre}
\end{align}
where $\{P_\mu:\mu \in \mathbb N_0\}$ are the Legendre polynomials
\cite{olver}.  Each $a_\mu(X/\Delta)$ is an oscillatory function with
$\mu$ zeros within the subdiffraction region $|X| \le \Delta$.  To
illustrate, Fig.~\ref{legendre_plot} plots the monomials $\xi^\mu$ and
the orthogonal polynomials given by Eq.~(\ref{legendre}) up to
$\mu = 8$.

\subsection{\label{sec_spade}Moment measurement and Fourier analysis by spatial-mode
demultiplexing (SPADE)}
The measurement of object moments by SPADE has been extensively
studied in the context of quantum-inspired imaging
\cite{tsang17,tsang18a,tsang19,tsang19b,tsang21a,zhou19,yang16,bonsma19,liang21a}.
The relation between the moments and the Fourier coefficients given by
Eq.~(\ref{fourier_moment}) is a helpful insight in bridging the
literature on moment estimation to other areas of superresolution
research, as the concept of resolution in optics is commonly framed in
terms of Fourier analysis \cite{villiers,goodman}, and the relation
between resolution and the moments is less clear.  Here I review the
principle of SPADE and discuss how it can be used to learn the Fourier
coefficients.

Let $\psi(x)$ be the complex-valued point-spread function of a
diffraction-limited imaging system for the optical field, where
$x \in \mathbb R$ is the image-plane coordinate that is normalized
with respect to the magnification factor \cite{goodman} and the
function is normalized as $\intall |\psi(x)|^2 dx = 1$. The optical
transfer function, defined as
\begin{align}
\Psi(k) \equiv \frac{1}{\sqrt{2\pi}} \intall \psi(x) \exp(-ikx) dx,
\end{align}
is assumed to have a finite width; common examples in optics include
the Gaussian
\begin{align}
\Psi(k) &= \qty(\frac{2}{\pi})^{1/4}\exp(-k^2),
\label{gauss_Psi}
\end{align}
and the rectangle
\begin{align}
\Psi(k) = \frac{\iverson{|k| \le 1}}{\sqrt{2}}.
\label{rect_Psi}
\end{align}
With direct imaging, which measures the intensity on the image plane
with an image sensor, the expected image intensity is
proportional to \cite{goodman,goodman_stat}
\begin{align}
  f(x) &\equiv \intall |\psi(x-X)|^2 F(X) dX.
\label{f}
\end{align}
The SPADE scheme put forth, on the other hand, processes the
image-plane light by further photonics before photodetection. In the
scheme, the light is first demultiplexed in terms of the
point-spread-function-adapted (PAD) basis
$\{\phi_q(x): q \in \mathbb N_0\}$ \cite{rehacek17,tsang18a}, where
each PAD mode is defined as
\begin{align}
  \phi_q(x) &\equiv \frac{(-i)^q}{\sqrt{2\pi}}\intall \Psi(k) g_q(k) e^{ikx} dk
\end{align}
and $g_q(k)$ is an orthonormal polynomial with respect to the
$|\Psi|^2$-weighted inner product
\begin{align}
\Avg{g_q(k),g_p(k)}_{|\Psi(k)|^2} \equiv \intall g_q^*(k) g_p(k)|\Psi(k)|^2 dk =
\delta_{qp}.
\end{align}
For example, if $\Psi(k)$ is the Gaussian given by
Eq.~(\ref{gauss_Psi}), then
\begin{align}
g_q(k) &= \frac{1}{\sqrt{q!}}\He_q(2k),
\end{align}
where $\{\He_q:q \in \mathbb N_0\}$ are the probabilist's Hermite
polynomials \cite{olver}, or if $\Psi(k)$ is the rectangle given by
Eq.~(\ref{rect_Psi}), then
\begin{align}
  g_q(k) = \sqrt{2q+1}P_q(k).
\label{legendre_g}
\end{align}
The orthonormality of the orthogonal polynomials implies the
orthonormality of the PAD basis with respect to the inner product for
the optical fields, viz.,
\begin{align}
\Avg{\phi_q,\phi_p} &\equiv 
\intall \phi_q^*(x) \phi_p(x) dx = \Avg{g_q,g_p}_{|\Psi|^2} = \delta_{qp}.
\end{align}
As the image-plane optical fields are spanned by
\begin{align}
\qty{\psi_X: \psi_X(x) = \psi(x-X), |X| \le \Delta},
\end{align}
the identity 
\begin{align}
\sum_{q=0}^\infty \abs{\Avg{\phi_q,\psi_X}}^2 &= 1
\quad
\forall |X| \le \Delta
\end{align}
implies that all the image-plane photons can be captured in the PAD
basis and the demultiplexer can be lossless, at least in principle.
The identity can be proved for any $\Psi$ that satisfies
$\intall \exp(c|k|) |\Psi(k)|^2 dk < \infty$ for some $c > 0$ by
writing
\begin{align}
\Avg{\phi_q,\psi_X} = i^q \Avg{g_q(k),e^{-ikX}}_{|\Psi(k)|^2},
\label{transit_amp}
\end{align}
noting that the orthogonal polynomials $\{g_q\}$ are complete in the
$L_2(|\Psi|^2)$ space for the assumed class of $\Psi$, and applying
Parseval's identity \cite{parth05}.

Although only a finite number of modes may be demultiplexed in
practice, physical intuition suggests that the coupling efficiency
from a subdiffraction object to a PAD mode should decrease rapidly
with increasing mode order, and a truncation of the demultiplexed
modes should not have a major impact on the total photon-collection
efficiency.

\begin{figure}[htbp!]
\centerline{\includegraphics[width=0.48\textwidth]{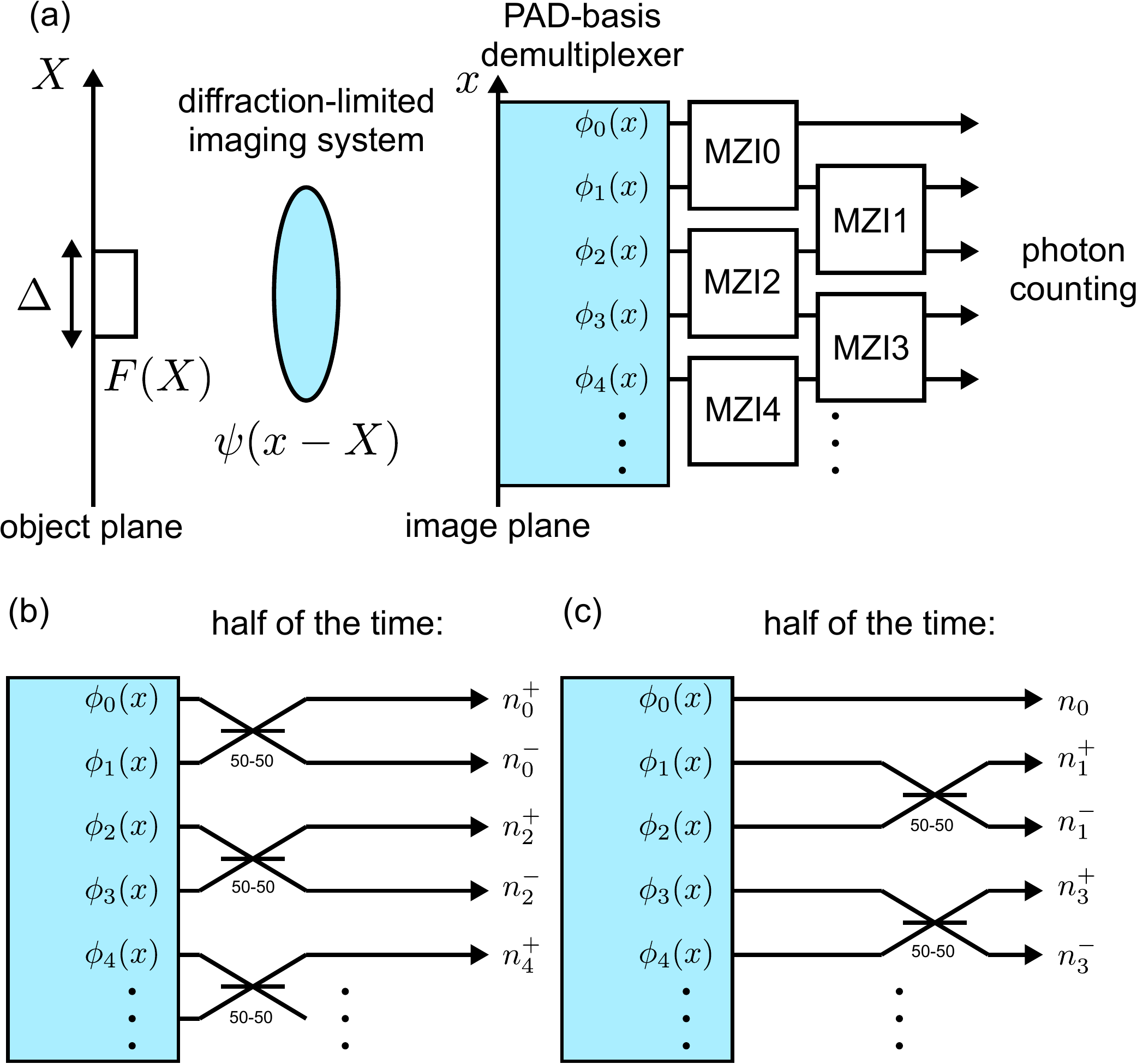}}
\caption{\label{scheme} The proposed SPADE scheme to measure the
  object moments and the Fourier coefficients for a distribution of
  incoherent sources.  (a) Basic setup, where $F(X)$ is the object
  intensity function, $\Delta$ is the object width, and $\psi(x-X)$ is
  the point-spread function of the imaging system for the optical
  field. The demultiplexer sorts the image-plane light in terms of the
  point-spread-function-adapted (PAD) basis $\{\phi_q(x)\}$, and each
  Mach-Zehnder interferometer (MZI) further combines a pair of the
  outputs. \new{Each MZI operates as a variable beamsplitter by having
    a phase modulator that controls the relative phase between the two
    arms.}  (b) Configuration of the MZIs for half of the observation
  time: MZI0, MZI2, ... should be set as 50-50 beamsplitters, while
  MZI1, MZI3, ... should be set as passthroughs. (c) Configuration for
  the other half of the time: MZI0, MZI2, ... should be set as
  passthroughs, while MZI1, MZI3, ... should be set as 50-50
  beamsplitters.  $n_q^\pm$ and $n_0$ are the photon counts that
  should be further processed to produce estimates of the moments and
  the Fourier coefficients. The expected photon counts are given by
  Eqs.~(\ref{n_mean})--(\ref{transit_prob}), while the estimators for
  the Fourier coefficients and the moments are given by
  Eqs.~(\ref{beta_est}) and (\ref{theta_est}).}
\end{figure}

After the demultiplexer, the light passes through pairwise
interferometers before photon counting, as depicted in
Fig.~\ref{scheme}. The expected value ($\expect$) of each photon count
is
\begin{align}
\expect\qty(n_q^\pm) &= N f_q^\pm,
\quad
q  \in \mathbb N_0,
\label{n_mean}
\\
f_q^\pm &\equiv \frac{1}{2} \intall h_q^\pm (X) F(X) dX,
\label{fq}
\\
h_q^\pm(X) &\equiv 
\abs{\Avg{\frac{\phi_q\pm \phi_{q+1}}{\sqrt{2}},\psi_X}}^2,
\label{transit_prob}
\end{align}
where $N$ is the expected photon number detected in all outputs and
$h_q^\pm(X)$ is a transition probability of each photon reaching an
output as a function of the point-source displacement $X$.  Note that
all the $h_q^\pm(X)$ functions are bandlimited because of the
bandlimited $\psi_X$ and $\phi_q$, so each $\expect(n_q^\pm)$ is, in
effect, the outcome of passing $F(X)$ through a linear bandlimited
filter.

The setup enables both even and odd moments to be measured; without
the interferometers, only even moments can be measured
\cite{tsang18a}.  To see this, note that, since
$\exp(-ikX) = \sum_p (-ikX)^p/p!$ and
$\avg{g_q(k),k^p}_{|\Psi(k)|^2} = 0$ for $p < q$, the transition
amplitude for each demultiplexer output given by
Eq.~(\ref{transit_amp}) becomes
\begin{align}
\Avg{\phi_q,\psi_X} &\sim H_{q} X^q -i H_{q}'X^{q+1},
\label{transit_amp2}
\end{align}
where $H_{q} \equiv \avg{g_q(k),k^q}_{|\Psi(k)|^2}/q!$ and
$H_q' \equiv \avg{g_q(k),k^{q+1}}_{|\Psi(k)|^2}/(q+1)!$ are real constants
and $\sim$ means identical in the leading order for
$|X| \le \Delta \ll 1$. The transition probability
given by Eq.~(\ref{transit_prob}) becomes
\begin{align}
h_q^\pm(X) &\sim \frac{H_q^2}{2} X^{2q}\pm H_{q} H_{q+1} X^{2q+1},
\end{align}
the expected photon count becomes
\begin{align}
\expect\qty(n_q^\pm) &\sim 
\frac{N}{2} \qty(\frac{H_q^2}{2}\theta_{2q} \pm H_{q} H_{q+1} \theta_{2q+1}),
\label{output}
\end{align}
where each $\theta_\nu$ is an object moment defined by
Eq.~(\ref{moment}), and the addition and subtraction of the photon
counts give
\begin{align}
\expect\qty(n_q^+ + n_q^-)
&\sim \frac{N H_q^2}{2}\theta_{2q},
\label{output_plus}
\\
\expect\qty(n_q^+ - n_q^-)
&\sim N H_q H_{q+1}\theta_{2q+1}.
\label{output_minus}
\end{align}
Thus, the outputs $n_q^+\pm n_q^-$ can be used to estimate the moments
$\theta_{2q}$ and $\theta_{2q+1}$, and the moment estimates can then
be plugged into Eq.~(\ref{fourier_moment}) to estimate the Fourier
coefficients. 

\begin{figure}[htbp!]
\centerline{\includegraphics[width=0.48\textwidth]{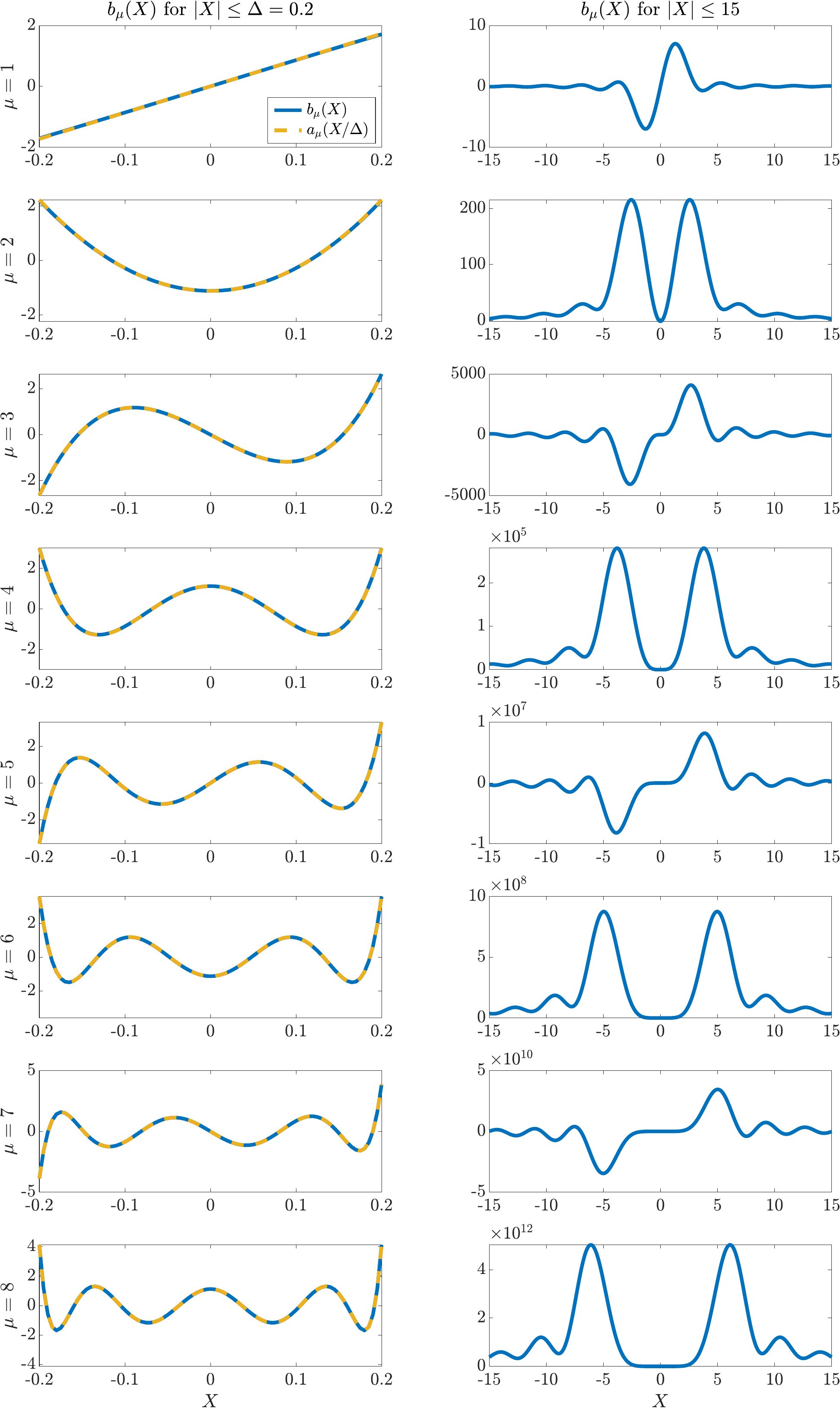}}
\caption{\label{superosc}Left column: plots of the exact filter
  function $b_\mu(X)$ (solid blue lines) from $\mu = 1$ to $\mu = 8$,
  as defined by Eqs.~(\ref{beta_est})--(\ref{filter}) and implemented
  by the measurement in Fig.~\ref{scheme}, with the estimators given
  by Eqs.~(\ref{beta_est}) and (\ref{theta_est}).  The functions are
  seen to be close to the desired filters $a_\mu(X/\Delta)$ (dash
  yellow lines). Right column: plots of $b_\mu(X)$ for a wider range
  of $X$, showing the large sidelobes.  The plots assume that
  $\Delta = 0.2$, $R(\xi)$ is the rectangle function given by
  Eq.~(\ref{rect}), $\{a_\mu(\xi)\}$ are the Legendre polynomials
  given by Eq.~(\ref{legendre}), $\Psi(k)$ is the rectangle function
  given by Eq.~(\ref{rect_Psi}), $\{g_q(k)\}$ are the Legendre
  polynomials given by Eq.~(\ref{legendre_g}), and the $b_\mu$
  functions are computed using Eqs.~(\ref{beta_est}),
  (\ref{theta_est}), and (\ref{n_mean})--(\ref{transit_prob}).
  \new{To compare the feature sizes here with the diffraction limit,
    note that direct imaging according to Eq.~(\ref{f}) completely
    blocks any sinusoidal component $\exp(i\kappa X)$ of the input
    function $F(X)$ when $\mathbb R$ is the function domain and
    $|\kappa| \ge 2$, so a diffraction-limited feature size can be
    defined as the period $2\pi/|\kappa| = \pi$, which is much larger
    than the feature sizes of the oscillations shown in the left
    column.}}
\end{figure}

To be more precise about the counting statistics, assume that the
photon counts are independent and Poisson random variables, which are
an excellent assumption for natural or fluorescent sources at optical
frequencies \cite{tsang19a,pawley,goodman_stat,zmuidzinas03}.  Assume
also that $N$ is unknown, for generality. Let the total photon number
detected in all outputs be $L$.  Conditioned on $L$, the count
statistics become multinomial.  Then an estimator of $\beta_\mu$ can
be constructed as follows:
\begin{align}
\check\beta_\mu &= \sum_{\nu=0}^\mu \frac{A_{\mu\nu}}{\Delta^\nu}
\check\theta_\nu,
\label{beta_est}
\\
\check\theta_\nu &= 
\begin{cases}
1, & \nu = 0,\\
(n_q^+-n_q^-)/(L H_qH_{q+1}),& \nu \textrm{ odd}, \nu = 2q+1,\\
2(n_q^++n_q^-)/(L H_q^2) , & \nu \textrm{ even}, \nu = 2q.
\end{cases}
\label{theta_est}
\end{align}
For a given $L$,
$\expect(n_q^\pm) = L f_q^\pm = L \intall h_q^\pm(X) F(X) dX$, and the
expected value of $\check\beta_\mu$ can be expressed as
\begin{align}
\expect\qty(\check\beta_\mu) &= \intall b_\mu(X) F(X) dX,
\label{filter}
\end{align}
where $b_\mu(X)$ is a filter function that should approximate the
oscillatory $a_\mu(X/\Delta)$ in the region $|X| \le \Delta$.  As
$b_\mu(X)$ is a linear combination of the bandlimited $h_q^\pm(X)$
functions, it should be a superoscillatory function with large
sidelobes; Fig.~\ref{superosc} supports this assertion by plotting
$b_\mu(X)$ for $\mu = 1,2,\dots,8$.  The filter functions also
resemble the prolate-spheroidal functions that are well known in
superresolution theory \cite{villiers}.

Note that each $\check\beta_\mu$ involves PAD modes up to order
$q = \ceil{\mu/2}$ only, so the estimation of each Fourier coefficient
requires the demultiplexing of only a finite number of modes.  The
efficiency of a subdiffraction object coupling into a PAD mode is
expected to go down for higher mode orders, so there are only a finite
number of modes that offer useful SNRs and a finite number of Fourier
coefficients that can be estimated accurately in practice.

\new{Note also that the moment estimator given by
  Eq.~(\ref{theta_est}) is based on a Taylor-series approximation made
  in Eqs.~(\ref{transit_amp2})--(\ref{output_minus}), so the estimator
  is unbiased only in the leading order
  ($\expect(\check\theta_\nu) \sim \theta_\nu$).  Since the $b_\mu(X)$
  filter function defined by Eqs.~(\ref{beta_est})--(\ref{filter}) is
  exactly equal to the orthogonal polynomial $a_\mu(X/\Delta)$ only
  when the moment estimator is exactly unbiased
  ($\expect(\check\theta_\mu) = \theta_\nu$), the implemented
  $b_\mu(X)$, as plotted in Fig.~\ref{superosc} for example, is only
  an approximation of $a_\mu(X/\Delta)$. The bias can be reduced by
  measuring more modes and using a more complicated moment estimator
  \cite{tsang19b}, although Fig.~\ref{superosc} shows that $b_\mu(X)$
  and $a_\mu(X/\Delta)$ can be very close to begin with and the
  $b_\mu(X)$ functions still exhibit the desired oscillatory
  behaviors.  }

\subsection{\label{sec_implement}Implementations of SPADE}
The PAD-basis demultiplexer is the key component in the proposed
scheme.  Also known as a mode sorter in other areas of optics, the
demultiplexer can be implemented by many different methods, as
reviewed in Refs.~\cite{tsang19a,piccardo21}. I mention only a few
implementations that have recently been demonstrated for the purpose
of superresolution:
\begin{enumerate}
\item Multi-plane light conversion (MPLC) \cite{boucher20}, which
  involves light propagation through a series of specially designed
  phase plates. \new{With this device, the outputs are well separated
    Gaussian beams and can be coupled into single-mode fibers, so the
    MZIs can be implemented using standard components. There also
    exist algorithms to design the phase plates for more general
    unitary operations \cite{morizur,fontaine19}.}

\item Super-resolved position localization by inversion of coherence
  along an edge (SPLICE) \cite{tham17,bonsma19}, which involves a
  binary phase plate and a single-mode fiber to perform mode
  selection.

\item Mode-selective up-conversion \cite{donohue18,zhang20a,ansari21},
  which involves an optical pump with an appropriate spatial or
  temporal profile to up-convert the desired mode of the input
  via sum frequency generation in a nonlinear medium.

\item Image-inversion interferometry \cite{sliver,tang16},
which involves a Mach-Zehnder interferometer that inverts
the spatial profile of the optical beam in one of the arms.

\item Mode-selective heterodyne detection \cite{yang16,pushkina21},
  which involves the interference of the input with a local oscillator
  with an appropriate spatial profile at an image sensor.
\end{enumerate}
It is beyond the scope of this work to discuss the experimental
details and the relative merits of different implementations. Here I
only emphasize the general fact that, even though SPADE is by no means
trivial to implement experimentally, it requires only known and
accessible optical materials and technologies.

It is interesting to note that coronagraphs \cite{guyon06} and nulling
interferometers \cite{serabyn19} in astronomy turn out to perform mode
sorting not unlike some simple versions of SPADE, although their
superiority to direct imaging in rigorous statistical terms, their
quantum optimality, and their applicability to more general imaging
problems do not seem to have been studied before. The practical
success of those instruments in astronomy offers encouragement that
more advanced indirect imaging methods such as SPADE should remain
viable in realistic conditions and in applications beyond astronomy.

\subsection{Possible generalizations}
\begin{figure}[htbp!]
\centerline{\includegraphics[width=0.48\textwidth]{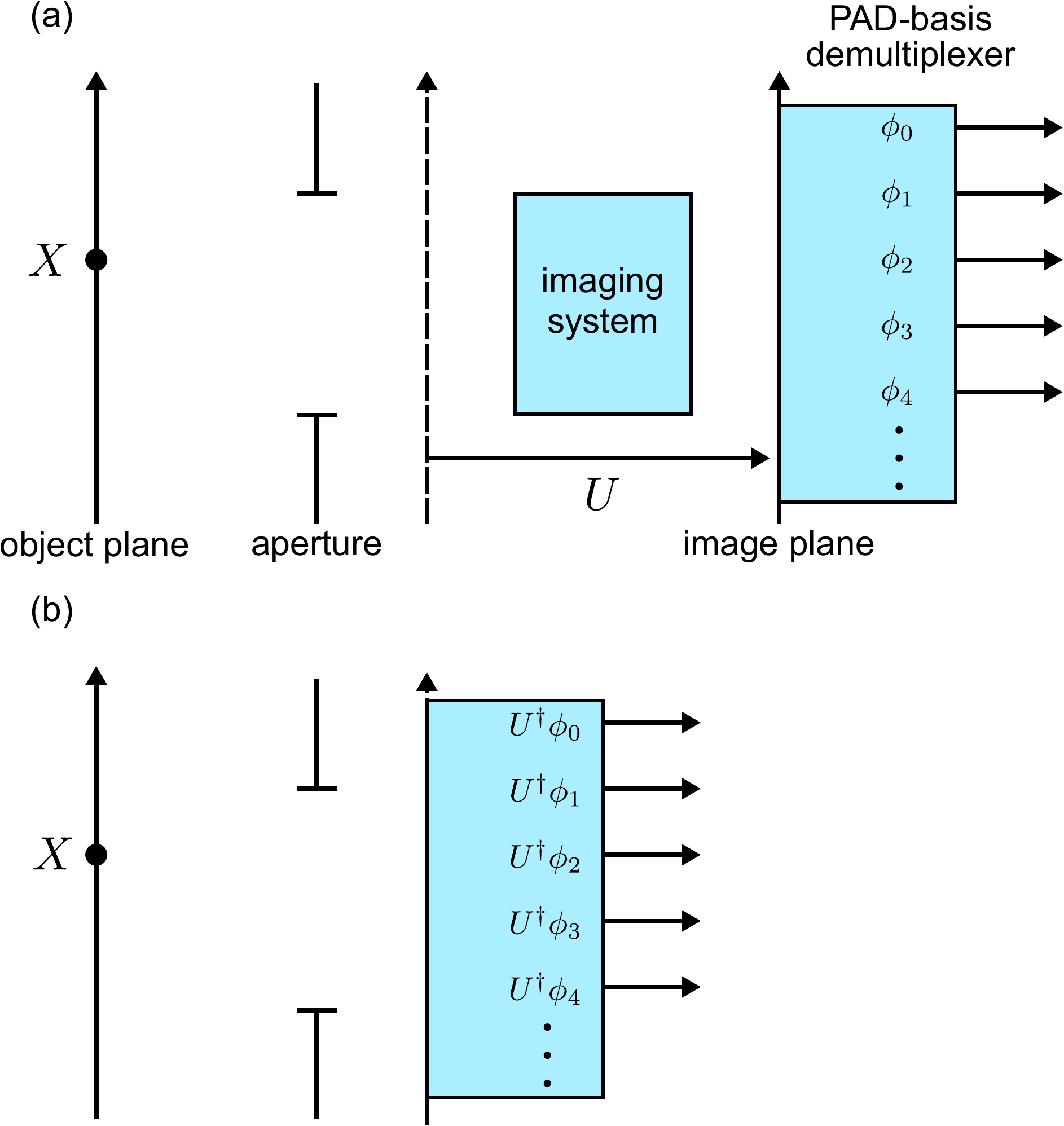}}
\caption{\label{planes}(a) In the original proposal, demultiplexing in
  terms of the PAD basis is applied to the image-plane light.  The
  optical field on the image plane can be related to the field on
  another plane after the aperture by a unitary operator $U$.  (b) If
  demultiplexing is applied on that plane, the same results in the
  original proposal can be reproduced by modifying the basis via the
  adjoint operator $U^\dagger$.}
\end{figure}

If the demultiplexer is not placed on an image plane, the
wavefunctions of the modes to be demultiplexed should be modified.
Let $\alpha_X(x)$ be the optical field on a certain plane after the
aperture produced by a point source with displacement $X$ and suppose
that the field on that plane is to be demultiplexed. In principle, the
image-plane $\psi_X$ is related to $\alpha_X$ by a unitary operator
$U$ that models the propagation through the optical components between
the two planes \cite{goodman}, viz.,
\begin{align}
\psi_X(x) &= (U \alpha_X)(x) = \intall U(x,x')\alpha_X(x') dx'.
\end{align}
Then the preceding theory still holds if the wavefunctions of the PAD
modes are modified by the adjoint operator $U^\dagger$ (which is also
unitary), such that
\begin{align}
\Avg{U^\dagger \phi_q,\alpha_X} = \Avg{\phi_q,U \alpha_X} =
\Avg{\phi_q,\psi_X},
\end{align}
and the transition amplitude for each demultiplexer output is still given
by Eq.~(\ref{transit_amp}). In other words, to compute the wavefunctions
of the modes to be demultiplexed, one simply backpropagates the
wavefunctions of the PAD modes from the image plane to the desired
input plane of the demultiplexer, as illustrated by Fig.~\ref{planes}.

The theory here can be applied to imaging with multiple apertures,
such as multi-telescope interferometry \cite{labeyrie}, by specifying
an optical transfer function $\Psi(k)$ that models the total aperture.
Each PAD mode is then a ``supermode'' that is a superposition of
optical fields from all the apertures, and the demultiplexer would
require a more elaborate optical processor that combines them
coherently.

The performance of the proposed scheme may be improved further by
optimizing the interferometry and the time allocation between the
interferometer settings, although Sec.~\ref{sec_quantum} later shows
that the present scheme is already quantum-optimal in terms of its
error scaling with the object size.

Regardless of the object size $\Delta$, unbiased estimators of the
moments, and therefore the Fourier coefficients, may still be
constructed from the outputs of the proposed SPADE scheme
\cite{tsang19a}.  If $\Delta$ is not subdiffraction, however, it
becomes unclear whether SPADE offers a statistical advantage over
direct imaging. Moreover, as the efficiency of coupling to a
higher-order mode becomes higher for a larger object, more modes will
acquire significant photon numbers and will need to be measured in
order to minimize the bias of the estimators.

Generalizations for imaging in two dimensions and spectroscopy are
also possible by following Refs.~\cite{tsang18a,ansari21,mazelanik22}.

\section{Statistical analysis}

\subsection{Error with SPADE}
Although the sidelobes of the filter functions are large for a
subdiffraction object size $\Delta$, they occur for larger $X$ and are
irrelevant to an object of subdiffraction size. Of more fundamental
concern is the estimation error.  Assume $\Delta \ll 1$ and the
asymptotic notations $O[p(\Delta)]$ (order at most $p(\Delta)$),
$\Omega[p(\Delta)]$ (order at least $p(\Delta)$), and
$\Theta[p(\Delta)]$ (order exactly $p(\Delta)$) for $\Delta \to 0$
\cite{knuth76}.  With multinomial statistics and
\begin{align}
f_q^\pm = \Theta(\Delta^{2q}),
\end{align}
the variance ($\vari$) of $\check\beta_\mu$ can be expressed as
\begin{align}
\vari\qty(\check\beta_\mu)
&\sim \frac{(A_{\mu\mu})^2}{\Delta^{2\mu}} \vari\qty(\check\theta_\mu)
= \frac{(A_{\mu\mu})^2}{\Delta^{2\mu}}  \frac{\Theta(\Delta^{2\floor{\mu/2}})}{L}
\nonumber\\
&= \frac{\Theta(\Delta^{-2\ceil{\mu/2}})}{L},
\label{var_beta}
\end{align}
Unfortunately---but perhaps unsurprisingly---the error gets worse for
smaller $\Delta$, especially for higher $\mu$.  The achievable
resolution enhancement, in terms of the number of accurately estimated
Fourier coefficients, depends on $\Delta$ and the photon budget
$N = \expect(L)$.

\subsection{\label{sec_crb}Comparison with the Cram\'er-Rao bounds (CRBs) for direct
  imaging}

The superiority of the proposed SPADE scheme over any processing of
the image-plane intensity can be proved by comparing the error of the
former with a Cram\'er-Rao bound (CRB) for direct imaging
\cite{tsang19a}. Ideal direct imaging can be modeled as a measurement
of a spatial Poisson process with the intensity function $Nf(x)$,
where $f(x)$ is given by Eq.~(\ref{f}) \cite{goodman_stat}.  The
parameter space is assumed here to be the set of all probability
densities for $F(X)$, while the Fourier coefficient $\beta_\mu[F]$
given by Eq.~(\ref{filter}) is taken as the parameter of interest.

If the point-spread function $|\psi(x)|^2$ is Gaussian, the exact CRB
for semiparametric moment estimation with direct imaging can be
derived using the techniques in
Refs.~\cite{tsang18a,tsang19b,bickel93}, despite the infinite
dimensionality of the parameter space.  Furthermore, there exists an
efficient unbiased estimator that attains the bound
\cite{tsang19b}. As each Fourier coefficient $\beta_\mu$ is a linear
combination of the moments, it is straightforward to derive the CRB
for each $\beta_\mu$ from the results in
Refs.~\cite{tsang18a,tsang19b}. Conditioned on the total photon number
$L$, the CRB is
\begin{align}
\textrm{MSE}^{(\textrm{direct})} \ge
\mathsf C^{(\textrm{direct})} &= \frac{1}{L}\qty[ B C^{-1} M \qty(C^{-1})^\top B^\top]_{\mu\mu}
\label{exact_CRB}
\\
&= \frac{\Theta(\Delta^{-2\mu})}{L},
\label{exact_CRB_order}
\end{align}
where $\textrm{MSE}^{(\textrm{direct})}$ is the mean-square error of
direct imaging with any unbiased estimator of $\beta_\mu$,
$M$ is the image moment matrix given by
\begin{align}
M_{uv} &\equiv \Avg{x^u,x^v}_{f(x)} - \Avg{x^u,1}_{f(x)} \Avg{x^v,1}_{f(x)},
&
u,v \in \mathbb N_0,
\end{align}
$B$ and $C$ are lower-triangular matrices given by
\begin{align}
C_{uv} &\equiv \iverson{u \ge v} \begin{pmatrix}u\\ v \end{pmatrix}
\intall x^{u-v} |\psi(x)|^2 dx,
\\
B_{uv} &\equiv \iverson{u \ge v} \frac{A_{uv}}{\Delta^v},
\label{B}
\end{align}
and $\top$ denotes the transpose. By virtue of the
lower-triangularity of $B$ and $C$, only the submatrices of $B$, $C$,
and $M$ with entries up to $u = v = \mu$ are needed to compute
$\mathsf C^{(\textrm{direct})}$.  A comparison of Eq.~(\ref{var_beta})
with Eq.~(\ref{exact_CRB_order}) suggests that SPADE is superior to
direct imaging for $\mu \ge 2$.

If $|\psi(x)|^2$ is not Gaussian, the exact semiparametric CRB is much
more difficult to derive.  An alternative CRB can be obtained via the
parametric-submodel approach in Ref.~\cite{tsang21a}. It involves the
judicious choice of an unfavorable parametric submodel
$F_\vartheta(X)$ with a scalar parameter $\vartheta \in \mathbb R$ and
$F_0(X)$ being set as the true object intensity.  \new{Then the CRB of
  the submodel, denoted as $\tilde{\mathsf C}^{(\textrm{direct})}$, is
  a lower bound on the exact CRB $\mathsf C^{(\textrm{direct})}$ for
  the semiparametric problem \cite{bickel93}, viz.,
\begin{align}
\mathsf C^{(\textrm{direct})}
\ge 
\tilde{\mathsf C}^{(\textrm{direct})}.
\label{submodel_bound}
\end{align}
This lower bound makes intuitive sense, as the submodel assumes fewer
unknown parameters and should therefore permit a lower uncertainty in
the parameter of interest.}  Conditioned on $L$, the submodel CRB can
be expressed as
\begin{align}
\tilde{\mathsf C}^{(\textrm{direct})} &= \frac{(\partial\beta_\mu)^2}{L J},
&
\partial\beta_\mu &\equiv \eval{\pdv{\beta_\mu[F_\vartheta]}{\vartheta}}_{\vartheta = 0},
\label{submodel_CRB}
\end{align}
where $\partial$ denotes the partial derivative at the true
$\vartheta = 0$ and $J$ is the per-photon Fisher information for the
submodel. \new{Since the parameter space consists of all probability
  densities, there is considerable freedom in specifying a submodel. A
  convenient one is \cite{bickel93}
\begin{align}
F_\vartheta(X) &= \frac{\{1 + \tanh[\vartheta c_\mu(X/\Delta)]\} F_0(X)}{\intall
(\textrm{numerator}) dX},
\label{submodel}
\end{align}
where $c_\mu$ satisfies the zero-mean condition
$\intall c_\mu(X/\Delta) F_0(X) dX = 0$ at the truth, such that
\begin{align}
\partial F_\vartheta(X) = c_\mu\qty(\frac{X}{\Delta}) F_0(X).
\label{submodel_score}
\end{align}
Since $F_0(X)$ is assumed to coincide with the true density and
$F_\vartheta(X)$ is a valid probability density for any
$\vartheta \in \mathbb R$, Eq.~(\ref{submodel}) satisfies the
requirements of a parametric submodel for any zero-mean $c_\mu$, and
the CRB computed from the submodel can be used in
Eq.~(\ref{submodel_bound}).  The choice of an unfavorable submodel
then boils down to the choice of $c_\mu$.} A fruitful choice of
$c_\mu$ made in Ref.~\cite{tsang21a} is an orthogonal polynomial with
respect to the true $W(\xi) = W_0(\xi)$, which is defined by
$F_0(X) = \Delta^{-1} W_0(X/\Delta)$ in the same manner as
Eq.~(\ref{W}), while $\mu$ is chosen to match that of $\beta_\mu$.
Then
\begin{align}
\partial \beta_\mu &= \intall a_\mu\qty(\frac{X}{\Delta})
c_\mu\qty(\frac{X}{\Delta}) F_0(X) dX
\\
&= A_{\mu\mu} \Avg{\xi^\mu,c_\mu(\xi)}_{W_0(\xi)} = \Theta(1).
\end{align}
The Fisher information, on the other hand, is given by
\begin{align}
J &= \intall \frac{[\partial f_\vartheta(x)]^2}{f_0(x)} dx,
\label{J}
\\
f_\vartheta(x) &\equiv \intall \abs{\psi(x-X)}^2 F_\vartheta(X) dX.
\label{f_submodel}
\end{align}
Although the submodel CRB is a valid lower bound on the MSE for the
semiparametric problem, it is difficult to know how much looser it is
than the exact bound.

\begin{figure}[htbp!]
\centerline{\includegraphics[width=0.48\textwidth]{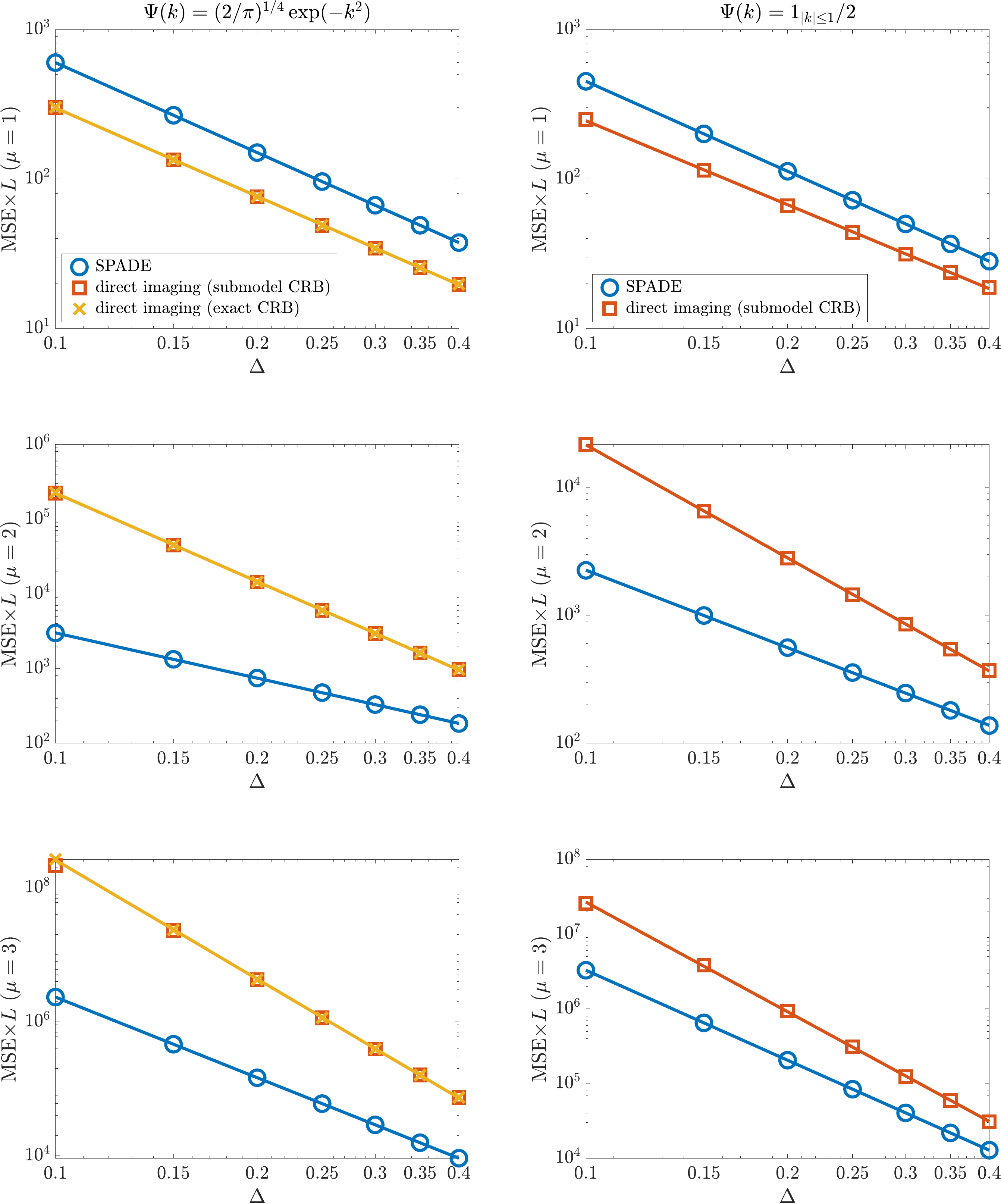}}
\caption{\label{superosc_errors}A numerical comparison of the SPADE
  errors with the CRBs for direct imaging in estimating the Fourier
  coefficients $\beta_\mu$, $\mu = 1,2,3$, assuming that the true
  object intensity is given by Eq.~(\ref{true_F}). In each plot of
  log-log scale, the vertical axis is the mean-square error (MSE)
  multiplied by the photon number $L$, while the horizontal axis is
  the object size $\Delta$ normalized with respect to the
  point-spread-function width. All the quantities are dimensionless by
  definition. The SPADE error is the variance of the estimator given
  by Eqs.~(\ref{beta_est}) and (\ref{theta_est}), assuming multinomial
  statistics for $\{n_q^\pm\}$ with a photodetection probability
  distribution given by Eqs.~(\ref{fq}) and (\ref{transit_prob}). For
  direct imaging, the exact semiparametric CRB is the
  $\mathsf C^{(\textrm{direct})}$ given by
  Eqs.~(\ref{exact_CRB})--(\ref{B}), while the submodel CRB is the
  $\tilde{\mathsf C}^{(\textrm{direct})}$ given by
  Eqs.~(\ref{submodel_CRB})--(\ref{f_submodel}) and computed
  numerically by discretizing the integrals in those equations. The
  left column assumes the Gaussian transfer function given by
  Eq.~(\ref{gauss_Psi}), while the right column assumes the rectangle
  $\Psi(k)$ given by Eq.~(\ref{rect_Psi}), in which case the exact CRB
  for direct imaging is unknown and only the submodel CRB is
  plotted. \new{The straight lines are least-squares linear fits of
    $\log\textrm{MSE}$ versus $\log \Delta$ and $\log\textrm{CRB}$
    versus $\log \Delta$ (assuming the exact CRB for the left column);
    their slopes differ from the corresponding theoretical exponents
    in Eqs.~(\ref{var_beta}), (\ref{exact_CRB_order}), and
    (\ref{submodel_CRB_order}) by at most a fractional error of 7\%
    only, where the fractional error is defined as
    $|\textrm{slope} - \textrm{theoretical exponent}|/
    |\textrm{theoretical exponent}|$.}  }
\end{figure}

Even with the submodel approach, the scaling of the CRB with respect
to $\Delta$ is difficult to derive if the point-spread function
$|\psi(x)|^2$ contains zeros \cite{tsang21a}.  I therefore resort to
numerics to compute the submodel bound here. The final numerical
results are plotted in Fig.~\ref{superosc_errors}. The figure plots
the variance of the estimator of $\beta_\mu$ ($\mu = 1,2,3$) with
SPADE and the CRBs for direct imaging against the object size $\Delta$
in log-log scale, assuming that the true object intensity is
\begin{align}
F(X) = F_0(X) = \frac{\iverson{|X|\le\Delta}}{2\Delta}.
\label{true_F}
\end{align}
The left column assumes a Gaussian optical transfer function with
$\Psi(k)$ given by Eq.~(\ref{gauss_Psi}), while the right column
assumes the rectangle function given by Eq.~(\ref{rect_Psi}). The
caption of Fig.~\ref{superosc_errors} contains further details about
the numerical analysis.

A comparison of the SPADE error given by Eq.~(\ref{var_beta}) and the
CRB given by Eq.~(\ref{exact_CRB_order}) suggests that SPADE has an
advantage only for $\mu \ge 2$, and indeed Fig.~\ref{superosc_errors}
supports this suggestion.  For $\mu = 1$ (first row of
Fig.~\ref{superosc_errors}), the direct-imaging CRBs are somewhat
lower than the SPADE error, while the scalings with respect to
$\Delta$ all roughly follow Eqs.~(\ref{var_beta}) and
(\ref{exact_CRB_order}). The gap may be partially attributed to the
fact that SPADE uses only half of the time to obtain the photon count
$n_0^-$ that contributes to $\check\theta_1$ and $\check\beta_1$.

For $\mu = 2,3$ (second and third rows of Fig.~\ref{superosc_errors}),
SPADE begins to show a substantial advantage. The scalings of the
SPADE error again follow Eq.~(\ref{var_beta}), while the scalings of
the direct-imaging CRBs follow Eq.~(\ref{exact_CRB_order}) for the
Gaussian $\Psi(k)$ (left column of Fig.~\ref{superosc_errors}). On a
side note, it is fortuitous here that the submodel CRB is so close to
the exact CRB for the Gaussian case, meaning that the chosen submodel
happens to be close to the least favorable submodel that gives the
exact CRB \cite{bickel93}.

For the rectangle $\Psi(k)$, the point-spread function
$|\psi(x)|^2 \propto \sinc^2 x$ contains zeros, which may enhance
the submodel Fisher information for direct imaging given by
Eq.~(\ref{J}) by a $\Theta(\Delta^{-1})$ factor \cite{paur18}.  The
plots of the submodel CRB in the right column of
Fig.~\ref{superosc_errors} and the arguments in Ref.~\cite{paur18}
motivate the conjecture that, when $|\psi(x)|^2$ contains zeros, the
submodel CRB for direct imaging obeys the scaling
\begin{align}
\tilde{\mathsf C}^{(\textrm{direct})}
&= 
\begin{dcases}
\frac{\Theta(\Delta^{-2})}{L}, &\mu = 1, \\
\frac{\Theta(\Delta^{-2\mu+1})}{L}, &\mu \ge 2,
\end{dcases}
\label{submodel_CRB_order}
\end{align}
although the tightness of this submodel bound for the semiparametric
problem remains unclear.

\subsection{\label{sec_quantum}Quantum Cram\'er-Rao bounds (QCRBs)}
The parametric-submodel approach can also be used to compute a quantum
CRB $\tilde{\mathsf H}$ that is valid for any measurement
\cite{tsang21a}. The result is
\begin{align}
\textrm{MSE} \ge \mathsf C \ge \mathsf H \ge
\tilde{\mathsf H} &= \frac{(\partial\beta_\mu)^2}{N K}  = 
\frac{\Omega(\Delta^{-2\ceil{\mu/2}})}{N},
\end{align}
where $\textrm{MSE}$ is now the mean-square error of any unbiased
estimator, $\mathsf C$ is the classical CRB for any measurement of the
image-plane light, $\mathsf H$ is the exact QCRB for the
semiparametric problem \cite{tsang20}, and $\tilde{\mathsf H}$ is the
QCRB for the submodel. The scaling
\begin{align}
K = O\qty(\Delta^{2\ceil{\mu/2}})
\end{align}
of the Helstrom information $K$ \cite{helstrom} for the submodel is
proved in Ref.~\cite{tsang21a}. With the detected photon number $L$
being close to the expected value $N$ for large $N$, the SPADE error
given by Eq.~(\ref{var_beta}) is quantum-optimal in terms of its
scaling with the object size.  

It is noteworthy that the quantum bound is invariant to any unitary
operation on the image-plane light, so the bound is applicable to more
general indirect imaging systems, such as stellar interferometry, that
are diffraction-limited but do not necessarily have a physical image
plane in its setup.

\section{\label{sec_practical}Practical concerns and open problems}
It is important to note that the theory here assumes ideal conditions
for both SPADE and direct imaging.  For SPADE, perfect demultiplexing
in the PAD basis and perfect interferometry are assumed, while for
direct imaging, an infinitesimal pixel size, an infinite number of
pixels with no gaps inbetween, and no interpixel crosstalk are
assumed.  For both methods, the object is assumed to have a strict
subdiffraction size, and no excess noise other than photon shot noise
are assumed in the photodetection.

A fair comparison of SPADE and direct imaging under practical
conditions is difficult to perform in theory, as they involve very
different optics and may be affected by current technical limitations
in very different ways. Here I only highlight one practical concern
for SPADE, which is crosstalk among the channels \cite{gessner20}. An
intuitive reason for the superiority of SPADE is that the filtering of
the lower-order modes from each channel reduces the output variance by
reducing the expected photon count given by Eq.~(\ref{output}) while
retaining some sensitivity to certain moments. If the $q$th output of
the PAD-basis demultiplexer is contaminated by some lower-order modes,
the transition amplitude given by Eq.~(\ref{transit_amp}) would
contain lower-order monomials ($\propto X^p$ with $p < q$), which in
turn introduce lower-order moments to the expected photon count given
by Eq.~(\ref{output}). The bias due to these parasitic terms may be
corrected if the lower-order modes and therefore the lower-order
moments are also measured.  Of more fundamental concern is the
increase in the variance that is also proportional to the expected
photon count. This variance increase leads to a suboptimal error
scaling with $\Delta$ for small $\Delta$.

Direct imaging under current technology is, of course, not perfect
either, and a useful comparison between the two methods under
practical conditions is perhaps best performed with experiments.

On the theoretical side, the exact quantum optimality of the proposal
here for the semiparametric problem remains an open question.  \new{On
  one hand, a more exact computation of the quantum bound than the one
  discussed in Sec.~\ref{sec_quantum} is needed and not at all trivial
  for the semiparametric problem because of the infinite
  dimensionality \cite{tsang21a,tsang20}.  On the other hand, it may
  be possible to optimize the scheme further by more complicated
  interferometry.}

Another interesting open problem is density estimation: the
reconstruction of the object intensity via the moments or the Fourier
coefficients, taking into account the positivity of the intensity and
any other prior information.  More advanced statistical methods will
be required to study the estimators, the performances, and the limits
\cite{tsybakov}.

\new{Finally, I should mention that the paraxial approximation and the
  assumption of incoherent light from the object are idealizations and
  may introduce systematic errors in practice. Given the small
  numerical aperture (N.A.)  in astronomy and remote sensing, there is
  no reason to doubt the accuracy of the paraxial approximation there,
  but one may need the full electromagnetic field theory to accurately
  model high-N.A.\ microscopy
  \cite{mortensen10,localization,zhang21}. As the concept of spatial
  modes remains valid in the full field theory \cite{fabre20}, a
  generalization of SPADE for high-N.A.\ imaging would be complicated
  but possible.

  There is also some recent academic interest in partially coherent
  sources \cite{liang21a,larson18,hradil21,kurdzialek22,tsang21}. The
  optical fields from any object must have a nonzero coherence length
  in principle \cite{kurdzialek22,born_wolf}. The coherence length of
  astronomical sources should be on the order of the wavelength
  \cite{born_wolf}, which is so much smaller than all the other length
  scales in the imaging problem that it is unlikely to make any
  noticeable difference in astronomy. Although partial coherence may
  be more important to sensing and microscopy, there is a dearth of
  experiments characterizing the coherence in those applications,
  making it difficult to even write down a realistic partially
  coherent model, let alone perform a useful analysis. Until a better
  model emerges, the incoherent model remains the gold standard in
  both astronomy \cite{goodman_stat,zmuidzinas03} and fluorescence
  microscopy \cite{pawley}.}

\section{Conclusion}
I have proposed a superoscillation measurement scheme for incoherent
imaging that overcomes the key limitations of previous
superoscillation techniques, such as inefficiency and questionable
advantage over computational techniques. Provably superior to direct
imaging and close to the quantum limit, the scheme put forth is
efficient in terms of both photon collection and statistical
performance. The theory shares similarities with several
superresolution concepts, thus establishing a common foundation for
future superresolution research.

To be sure, an implementation of SPADE with high efficiency and
fidelity is not trivial in practice. Its fundamental superiority, the
importance of the applications, and the rapid experimental progress in
photonics \cite{boucher20,fontaine19,piccardo21,norris19} should,
nonetheless, offer encouragement for its further development.

\bibliographystyle{myIEEEtran}
\bibliography{efficient_superoscillation}


\end{document}